\documentclass[12pt]{amsart}
\usepackage{amsmath,amsthm,amssymb}
\newtheorem{thm}{Theorem}[section]
\newtheorem{prop}[thm]{Proposition}
\newtheorem{cor}[thm]{Corollary}
\newtheorem{lemma}[thm]{Lemma}
\newtheorem{Def}[thm]{Definition}
\theoremstyle{definition}
\newtheorem{notn}[thm]{Notation}
\newtheorem{remark}[thm]{Remark}
\numberwithin{equation}{section}

\newcommand\cf{\textit{cf.}}
\newcommand\Z{{\mathbb Z}}
\newcommand\Q{{\mathbb Q}}
\newcommand\R{{\mathbb R}}
\newcommand\C{{\mathbb C}}
\newcommand\F{{\mathbb F}}
\DeclareMathOperator{\Hom}{Hom}
\DeclareMathOperator{\rank}{rank}
\DeclareMathOperator{\SO}{SO}
\DeclareMathOperator{\Orth}{O}
\newcommand\from{:}
\newcommand\cross{^\times}
\let\cont=\subseteq
\let\comp=\circ
\let\del=\partial
\let\eps=\varepsilon
\let\normal=\vartriangleleft
\let\semi=\rtimes
\let\tensor=\otimes
\let\into=\hookrightarrow
\newcommand\isom{\xrightarrow\sim}
\newcommand\bkt[1]{\langle #1\rangle}
\newcommand\mbar{{\tilde m}}
\newcommand{\T}{{\mathcal T}}
\newcommand{\g}{\mathfrak{g}}

\begin{document}
\title[Group Cohomology and Quasicrystals]
		{Applications of Group Cohomology to the Classification of
		Fourier-Space Quasicrystals}
\author{Benji N. Fisher}
\address{Mathematics Department \\ Boston College \\ Chestnut Hill, MA 02467}
\email{benji@member.AMS.org}
\urladdr{http://www2.bc.edu/$\sim$fisherbb}
\author{David A. Rabson}
\address{University of South Florida}
\email{davidra@ewald.cas.usf.edu}
\date{27 May, 2003}


\begin{abstract}
In 1962, Bienenstock and Ewald described the classification of crystalline
space groups algebraically in the dual, or Fourier, space.  Recently, the
method has been applied to quasicrystals and modulated crystals.
This paper phrases Bienenstock and Ewald's definitions in terms of group
cohomology.
A \textit{Fourier quasicrystal} is defined, along with its space group,
without requiring that it come from a quasicrystal in real (direct) space.
A certain cohomology group classifies the space groups associated to a given
point group and quasilattice, and the dual homology
group gives all gauge invariants.
This duality is exploited to prove several results that were previously known
only in special cases, including the classification of space groups for
quasilattices of arbitrary rank in two dimensions.
Extinctions in X-ray diffraction patterns and degeneracy of
electronic levels are interpreted as physical manifestations of non-zero
homology classes.
\vskip\baselineskip
\noindent{\tt PACS 2003: 02.10.Hh, 61.50.Ah, 61.44.Br, 61.44.Fw}
\end{abstract}
\maketitle

\setcounter{section}{0}
\addtocounter{section}{-1}
\section{Introduction} \label{sec:intro}

\subsection{Background}

The Penrose tilings of the plane~\cite{Penrose} have long-range order and
are very symmetrical, but they are not periodic.  A few years after the
discovery of these tilings, physical quasicrystals were discovered
\cite{Shechtman84,Janot94,Divincenzo99}.
These are solids with aperiodic structures that still have
long-range order and interesting symmetries, if one looks
in Fourier space.  In fact, the X-ray diffraction
patterns of some quasicrystals have five-fold symmetry, which is impossible
for periodic crystals.  Several mathematical models for quasicrystals have
been proposed.  A central question is how to classify the possible symmetries
of a quasicrystal, analogously to the classification of crystallographic groups,
which describe the symmetries of periodic crystals.

One approach to crystallography starts with the group~$\mathcal T$ of
translational symmetries of a crystal.  If the crystal is periodic, then
$\mathcal T$ is a lattice in~$\R^3$ (``real space'' or ``direct space'').  The
space group~$\mathcal G$ is the group of all isometries that preserve the
crystal, and it contains~$\mathcal T$ as a normal, Abelian subgroup.  The
quotient
$ G = \mathcal G / \mathcal T $
is called the point group of the crystal, and it can be considered a subgroup
of the orthogonal group~$\Orth(3)$.
A quasicrystal may have no translational symmetries, so this approach does not
generalize directly.  Instead, one can model a quasicrystal as the projection
into~$\R^3$ of a periodic crystal in a higher-dimensional space
(``superspace'')~\cite{JJdW}.

This paper takes a different approach to studying quasicrystals.
In 1962,
Bienenstock and Ewald \cite{B-E} introduced the ``Fourier-space approach'' to
classifying symmetries of crystals.
In this picture, a crystal is described by a periodic (electron or mass)
density function on~$\R^3$.
The Fourier coefficients of this density function are thus defined on the dual
lattice in the dual space~$\R^{3*}$ (``Fourier space'' or ``momentum space'').
The symmetries of the crystal can then be described in terms of these Fourier
coefficients, as discussed in Section~\ref{sec:definitions} below.
This approach has been applied to quasicrystals and modulated crystals
\cite{RWM88b,RWM,Rabson91,Mermin92a,Lifshitz94a,Lifshitz94b,Lifshitz97,K-M:1997,K-M:1999};
the only change is that one must relax the condition that the Fourier
coefficients be defined on a (discrete) lattice in Fourier space.

The Fourier-space approach to crystallography had
not been expressed explicitly
in terms of group cohomology until \cite{RabsonFisher02},
although the correspondence was
pointed out by Mermin~\cite{Mermin92a} and by Piunikhin~\cite{Piun-relation}.
(The direct-space approach has
been expressed in this language by Ascher, Janner, and others:
\cite{AJ-I,AJ-II,Schw,JJ,Hiller}.)
The goal of this paper is to describe the Fourier-space approach in terms of
group cohomology and to show how to take advantage of this well developed
theory.
Using this language, it is easy to prove and generalize results that other
authors have obtained, often by laborious calculation.
(In~\cite{K-M:1997}, these calculations are relegated to an appendix,
and in~\cite{RWM}, the reader is encouraged to skip them.)
This paper gives several examples.  Of course, some readers will
still be skeptical that it is worth learning about \textit{cocycles} and
\textit{coboundaries}.  Such readers should be convinced by
Theorem~\ref{thm:2D}, which classifies the space groups corresponding to a
given point group in two dimensions and a Fourier lattice (or quasilattice) of
arbitrary rank.  Crystallographers interested in applying the
cohomological language to quasicrystals are also referred
to \cite{RabsonFisher02,RabsonFisher03}.

For the reader familiar with the cohomological language, this should provide
one more interesting application of several familiar definitions and theorems.
The reader familiar with crystallography in Fourier space will
recognize in Section~\ref{sec:definitions} a new set of names for
several familiar ideas.

Previous work based on the Bienenstock-Ewald Fourier-space
approach, but not using the cohomological language, has been limited to
computations with quasilattices having explicit generators.
Some results~\cite{Lifshitz94a,Lifshitz94b} applied to only a few specific
quasilattices at a time and others~\cite{Rabson91,Mermin92a} only to
quasilattices
equivalent to principal ideals in the ring of cyclotomic integers~\cite{MRW46}.
Very little was known about quasilattices having non-minimal
ranks consistent with their rotational symmetry.
The techniques used in this paper
lift these restrictions.
The results presented here provide the
theoretical framework for the first complete classifications of space groups
in two and three dimensions~\cite{fisherrabsoninprep}.

\subsection{Summary of Results}

The first three sections describe the ideas studied in this paper:  Fourier
quasicrystals, their space groups, and their classification.
For the most part, we follow the
definitions and notation of Dr\"ager and Mermin~\cite{Dr-M}.
A Fourier quasicrystal~$\hat \rho$ is defined as the
coefficients of a formal Fourier series
\begin{equation} \label{eqn:formalFS}
 \rho(x) = \sum_{k \in L} \hat \rho(k) e^{2\pi i k \cdot x} ,
\end{equation}
where $L$~is a quasilattice:  a finitely generated additive group that
spans~$\R^{d*}$ but is not necessarily discrete.
Briefly, one associates to~$\hat \rho$ a triple
$(G, L, \{\Phi\})$,
where $G$~is a subgroup of the orthogonal group~$\Orth(d)$,
$L$~is a quasilattice in~$\R^{d*}$ stable under~$G$, and
$\{\Phi\}$ is a cohomology class in~$H^1(G,\hat L)$,
$ \hat L = \Hom(L, \R/\Z) $.
The ``point group''~$G$ can be thought of as the group of macroscopic
symmetries of~$\hat \rho$, and the triple
$(G, L, \{\Phi\})$
describes all symmetries, so we call the triple the ``symmetry type''
of~$\hat \rho$.
Section~\ref{sec:definitions} gives these definitions in detail.
Section~\ref{sec:realqc} discusses these definitions from the point of view of
the function~$\rho$ defined by the series~\eqref{eqn:formalFS}, assuming that
the series converges absolutely.  This assumption is made to keep the analysis
simple; a more comprehensive treatment of the relation between Fourier
quasicrystals and functions~$\rho$ is beyond the scope of this paper.
Section~\ref{sec:realqc} also discusses the relation between this and other
models of quasicrystals and the classical definition of space groups and point
groups.
Section~\ref{sec:classify} explains a programme for classifying symmetry types
that can be summarized by the phrase, ``$G$~first, then~$L$.''
This provides a context for most of the results proved in the later sections.
Other classifications first consider all quasilattices~$L$ of a given rank,
and then consider what point groups~$G$ can be associated to these lattices.
The approach used here is to fix the finite group~$G$ and then study the
quasilattices symmetric under~$G$.

The beauty of this programme is that, to classify $d$-dimensional symmetry
types, there is no need to leave dimension~$d$.  If one takes the direct-space
approach, the (super)space group of a quasicrystal is naturally a
crystallographic group in a higher-dimensional superspace,
with attendant complications.
On the other hand,
previous work using the Fourier-space approach, such as~\cite{RWM},
concentrated too
early on explicit generators of the quasilattice, and this led to unnecessary
restrictions (such as
requiring the lattice to be described by a principal ideal).
Concentrating first on the group~$G$ makes it possible to
calculate~$H^1(G,\hat L)$ for quite general two-dimensional quasilattices~$L$.
The authors are working on a paper that completes this programme in
dimension~$2$ and
hope, in future work, to do the same for dimension~$3$.

Each of the remaining sections illustrates the usefulness of the cohomological
language by taking a standard result about group cohomology and applying it to
crystallography.  Many of the applications are already known,
although in less generality.  In effect, the literature of
Fourier-space crystallography has been re-inventing the theory of group
cohomology.

Perhaps the most significant result of the paper (even though it is a direct
consequence of a standard result) is
Theorem~\ref{thm:dual}, which states that the cohomology group
$ H^1(G, \hat L) $
is dual to the homology group
$ H_1(G, L) $.
There are two ways of thinking of this duality.  One states
that elements of $H_1$
describe functions on~$H^1$ and so constitute
``fundamental gauge invariants'' in the language
introduced in Section~\ref{sec:definitions}.  In other words, this homology
group classifies all
possible ``gauge-invariant linear combinations of phases,'' the simplest of
which have found physical manifestations.
The opposite point of
view thinks of a cohomology class, or a gauge-equivalence class of phase
functions, as a linear function on the finite group
$ H_1(G, L) $.
This homology group is simpler, both conceptually and
computationally, than the cohomology group.  In fact, so long as one works
with~$\hat L$, the Pontrjagin dual of the lattice in Fourier space, it is
unclear to what
extent one is really taking a Fourier-space approach.  By concentrating
on~$H_1(G,L)$, we commit ourselves to this approach.

In the superspace approach to crystallography, the space group~$\mathcal G$ is
an extension of~$G$ by~$\mathcal T$, so it is described by an element of
$ H^2(G, \mathcal T) $.
The two approaches are connected by making the identifications
$ \mathcal T = \Hom(L, \Z) $
and
$ L = \Hom(\T, \Z) $.
\relax From this point of view,
Theorem~\ref{thm:dual}
states that
$ H^2(G, \mathcal T) $
is dual to
$ H_1\bigl(G, \mathcal \Hom(T, \Z) \bigr) $;
see Remark~\ref{rem:super}.

Sections \ref{sec:cyclic}~and~\ref{sec:res-inf} describe how the
restriction-inflation sequence and the
simple form of (co)homology of cyclic groups make the computation of
$ H_1(G, L) $
in dimensions 2~and~3 a tractable
problem.  As an application, Theorem~\ref{thm:2D} gives a complete description
of this homology group in the two-dimensional case.
In physical terms, the result means that the only
two-dimensional, non-symmorphic space groups are those whose point groups are
dihedral, with cyclic subgroup of order $N=2^e$.
This generalizes, without all the computation,
results already known in the restricted cases of
quasilattices of minimal rank, corresponding to principal ideals.
This theorem is closely related to one of Piunikhin:  Remark~\ref{rem:Piun}
discusses this further.

Another important part of crystallography is describing the physical
consequences of symmetry.  Preliminary computations suggest that, in two and
three dimensions, any homology group
$ H_1(G, L) $
is generated by cycles of a few simple types.  If so, and if
$ \{\Phi\} \in H^1(G, \hat L) $
is non-trivial, then
$ \bkt{\Phi,c} \ne 0 $
where $c$~is one of these simple cycles.
Non-vanishing gauge invariants tend to have physical implications, as
described in Section~\ref{sec:products}.  One of these is described by
K\"onig and Mermin~\cite{K-M:1997}, who suggest an
approach that generalizes to quasicrystals some crystalline phenomena usually
explained in terms of representation theory.  Proposition~\ref{prop:KM}
hints at how these ideas can be simplified and generalized using cohomology.
Another subject the authors hope to consider in future work is to
describe physical phenomena associated to each of the simple cycles.

Finally, we mention Proposition~\ref{prop:torsion}, Corollary~\ref{cor:inv},
and Proposition~\ref{prop:inv}.
The first two are results that were known
only in cases where the gauge-equivalence
(cohomology) classes had been calculated explicitly, and the third is
a non-computational proof of the result in the
appendix of~\cite{K-M:1999}.

This paper attempts to describe crystallography using group cohomology in a
way that can be understood both by those familiar with crystallography and by
those familiar with cohomology.  The reader will judge how well it succeeds.
In~\cite{RabsonFisher02} the authors describe many of the same ideas
explicitly in terms of cocycles,
and in~\cite{RabsonFisher03} they review the connection between
crystallography and algebraic topology for those unfamiliar with
the nomenclature of homological algebra.

% LATEX bug: this definition of foobox MUST precede the subsection,
% or the section will get incorporated into the box (probably conflicting
% use of box registers).
\newbox\foobox\newbox\leftbox\newbox\rightbox
\setbox\foobox\vtop{\hsize0.35\hsize\noindent lattice of
direct-space or superspace translations}
\subsection{Notation}
{
\everycr{\noalign{\vskip0.2\baselineskip}}
\def\strutt{\vtop to1.15\dp\strutbox{\null}\vbox to1.15\ht\strutbox{\null}}
\setbox\leftbox\hbox{\vtop{\halign{#\quad\quad&\hfil #\cr
$d$&coboundary map\strutt\cr
$\partial$&boundary map\strutt\cr
$\rho$&formal sum of coefficients\strutt\cr
$\hat\rho$&Fourier quasicrystal $L\rightarrow\C$\strutt\cr
$\Phi$&phase function $G\rightarrow\hat L$\strutt\cr
$\chi$&gauge function $L \rightarrow \R/\Z$\strutt\cr
$G$&point group of $\hat\rho$\strutt\cr
$G_L$&point group (holohedry) of $L$\strutt\cr
}%
}%
}%
\setbox\rightbox\hbox{\vtop{\halign{#\quad\quad&\hfil #\cr
$\mathcal G$&space group of $\hat\rho$\strutt\cr
$L$&quasilattice\strutt\cr
$L'$&$\Hom(L,\Q/\Z)$\strutt\cr
$\hat L$&$\Hom(L,\R/\Z)$, dual to $L$\strutt\cr
$M^G$&$\{x\mid gx=x, \ g\in G\}$\strutt\cr
$M_G$&$M/\bkt{\{kg-k~|~k\in M,\ g\in G\}}$\strutt\cr
$N_g$&$1+g+\dots+g^{N-1}$ if $g^N = 1$\strutt\cr
$\mathcal T$&\vtop{\unvbox\foobox}\strutt\cr
}%
}%
}%
$$%
\vcenter{%
\noindent\unhbox\leftbox\hfill\unhbox\rightbox
}
$$
}

\section{Definitions} \label{sec:definitions}

This section defines Fourier quasicrystals and their symmetry types, the main
objects of study in this paper.
Unfortunately, the language of quasicrystals is far from being standardized.
Definition~\ref{def:ql} follows~\cite{Piun-survey}, but
a few papers use the term \textit{quasilattice} to describe something else,
frequently a discrete set (not always closed under addition)
of direct-space translations.  Others refer
to a quasilattice (in the sense used here) as a (generalized)
lattice.  This paper uses
\textit{quasicrystal} as the most general term, encompassing periodic
and aperiodic crystals; some authors use the phrase
(generalized) \textit{crystal} for this, reserving the term
\textit{quasicrystal} for a particular kind of \textit{aperiodic} crystal.

The rest of the definitions mostly follow Dr\"ager and Mermin~\cite{Dr-M}.
The term \textit{Fourier quasicrystal} was suggested by an anonymous referee.
Section~\ref{sec:realqc} explains these
definitions in terms of quasicrystals in real space, again
following~\cite{Dr-M}.

\begin{Def} \label{def:ql}
Let $W$~be a Euclidean space.
A \textbf{quasilattice} in~$W$
is a finitely generated, additive subgroup
$ L \cont W $
that spans~$W$.
If, in addition, $L$~is discrete, then it is a \textbf{lattice}.
\end{Def}

It is well known that a quasilattice is discrete (hence a lattice) if and only
if its rank is the same as the dimension of~$W$.  Since the quasilattice~$L$
is required to span~$W$, the inequality
$ \rank(L) \ge \dim(W) $
always holds.

\begin{Def} \label{def:Fqc}
Let $L$~be a quasilattice.  A \textbf{Fourier quasicrystal} on~$L$ is a
function
$ \hat \rho \from L \to \C $
such that $L$~is generated, as an Abelian group, by the values of~$k$ for
which
$ \hat \rho(k) \ne 0 $.
\end{Def}

The requirement that the support of~$\hat\rho$ should generate~$L$ should
be thought of as a condition on~$L$, not on~$\hat\rho$, since
an arbitrary complex-valued function~$\hat\rho$ on a quasilattice~$L_1$
will be a Fourier quasicrystal on the quasilattice~$L$ generated by
$ \{\,k\mid\hat\rho(k)\not=0\,\}\cont L_1$.
The requirement that a quasilattice be finitely generated
underlies the International Union of Crystallography's (1992)
definition \cite{IUCr92} of ``crystal,'' referring to ``essentially discrete''
support of $\hat\rho$~.

\begin{Def} \label{def:gauge}
Let $L$~be a quasilattice.  A \textbf{gauge function} on~$L$ is an element of
the Pontrjagin dual
\begin{equation} \label{eqn:Pontrjagin}
	\hat L = \Hom(L, \R/\Z) .
\end{equation}
Two Fourier quasicrystals
$\hat \rho_1$~and~$\hat \rho_2$ on~$L$
are \textbf{indistinguishable} if there is a gauge function
$ \chi \in \hat L $
such that
\begin{equation} \label{eqn:indist}
	\hat\rho_2(k) = e^{2\pi i \chi(k)} \hat \rho_1(k)
	\qquad(\forall k \in L) .
\end{equation}
\end{Def}

A Fourier quasicrystal on
$ L \cont W $
can be thought of as the formal Fourier series~\eqref{eqn:formalFS}
where $x$~is in the dual space of~$W$.
The motivation for these definitions comes from thinking
of~$\hat \rho$ as the Fourier transform of such a function.
If $x$~is in the ``real space''~$\R^d$ of column vectors, then the space
spanned by~$L$ should be thought of as the dual space, so from now on
identify~$W$ with the ``Fourier space''~$\R^{d*}$ of row
vectors.\footnote{%
These conventions are convenient for making the connection
between direct and reciprocal space and for invoking
well-known results in cohomology \cite{Brown,C-F}.  In most
other work in Fourier-space crystallography ({\it e.g.},
\cite{RWM,Dr-M,RabsonFisher02,RabsonFisher03}), an element
of Fourier space is thought of as a column vector with a left
group action.  As a consequence, some results here ({\it
e.g.}, \eqref{eqn:gcc}) will take slightly different,
but entirely equivalent, forms.
}
Note that the orthogonal group~$\Orth(d)$ acts naturally on the left on~$\R^d$
and on the right on~$\R^{d*}$.

This paper makes no attempt to characterize the functions~$\rho$ for which a
series~\eqref{eqn:formalFS} can be defined, but Section~\ref{sec:realqc}
explains, under restrictive analytic assumptions, what indistinguishability
means in terms of the function~$\rho$ on real space.  A symmetry of a Fourier
quasicrystal is defined in terms of indistinguishability:

\begin{Def} \label{def:symmetry}
Let $L$~be a quasilattice in~$\R^{d*}$ and let $\hat \rho$ be a Fourier
quasicrystal on~$L$.
The \textbf{holohedry group}~$G_L$ is the subgroup of the orthogonal
group~$\Orth(d)$ consisting of all~$g$ such that
$ L \cdot g = L $.
A \textbf{symmetry} of~$\hat \rho$ is an element
$ g \in G_L $
such that $\hat\rho \comp g$ is indistinguishable from~$\hat\rho$.
In other words,
there is a gauge function
$ \Phi_g \in \hat L $
such that
\begin{equation} \label{eqn:symmetry}
	\hat \rho(k g) = e^{2\pi i \Phi_g(k)} \hat \rho(k)
	\qquad(\forall k \in L) .
\end{equation}
The \textbf{point group} of~$\hat \rho$ is the group~$G$ of all such symmetries.
The map
$ \Phi \from G \to \hat L $
is called a \textbf{phase function}.
\end{Def}

Sometimes the gauge functions~$\Phi_g$
are also called phase functions, but we avoid this usage.
Note that, since $\hat \rho$~is required to be non-zero on a set of generators
of~$L$ and $\Phi_g$~is linear on~$L$, the relation~\eqref{eqn:symmetry}
determines
$ \Phi_g(k) \in \R/\Z $
for all
$ k \in L $.
It is shown in \cite[\S1.2]{Piun-survey} that, even in dimension
$ d = 2 $,
a quasilattice may be symmetric under a rotation of infinite order, so
the holohedry group~$G_L$ is not always finite.  This paper usually assumes
that the point group~$\hat \rho$ is finite, but most results apply generally
to any finite subgroup of the point group.

The condition
$ k (gh) = (k g) h $
leads to the \textbf{group-compatibility condition}:
\begin{equation} \label{eqn:gcc}
	\Phi_{gh}(k) = \Phi_h(k g) + \Phi_g(k) .
\end{equation}
The natural right action of~$\Orth(d)$ on~$\R^{d*}$ induces a left action
of~$G$ on~$\hat L$.  In terms of this action, \eqref{eqn:gcc}~reads
$ \Phi_{gh} = g \Phi_h + \Phi_g \,$.
In other words,
$ \Phi \from G \to \hat L $
is a \textbf{cocycle} in~$Z^1(G, \hat L)$.

Now, let $\hat\rho_1$~and~$\hat\rho_2$ be
indistinguishable Fourier quasicrystals and let
$\chi$ be a gauge function as in~\eqref{eqn:indist}.  Then $\hat\rho_2$ has the
same point group as~$\hat\rho_1$, as can be seen by defining
\begin{equation} \label{eqn:gauge}
	\Phi^{(2)}_g(k) = \Phi_g(k) + \chi(k g - k) .
\end{equation}
The equation~\eqref{eqn:gauge} is called a \textbf{gauge equivalence}.
In terms of the left action of~$G$ on~$\hat L$, it reads
$ \Phi^{(2)}_g - \Phi_g = g \chi - \chi $,
which means that the difference of the two cocycles is the
\textbf{coboundary}, or \textbf{gauge transformation}, $g\chi-\chi$.
Since cohomologous cocycles (gauge-equivalent phase functions) express the
same symmetry of indistinguishable quasicrystals, it is
natural to associate to~$\hat \rho$ (or to its equivalence class under
indistinguishability) the cohomology class
$ \{\Phi\} \in H^1(G, \hat L) $.

\begin{Def} \label{def:symmetrytype}
Let $L$~be a quasilattice in~$\R^{d*}$ and let $\hat \rho$ be a Fourier
quasicrystal on~$L$.
The \textbf{symmetry type} of~$\hat \rho$ is the triple
$ (G, L, \{\Phi\}) $,
where $G$~is the point group of~$\hat \rho$ and
$\{\Phi\}$~is the cohomology class described above.
The \textbf{space group} of~$\hat \rho$ is the extension of~$G$ by
$ \Hom(L, \Z) $
corresponding to this cohomology class, as described in~\S\ref{sec:realqc}
below.
If the cohomology class is trivial,
then $\hat \rho$, or its space group, is called \textbf{symmorphic}.
\end{Def}

We use the term \textit{space group} even if $d=2$, where some authors might
prefer \textit{plane group}.
Section~\ref{sec:realqc} describes the space group from the real-space point
of view.
The symmorphic space group is simply the semidirect product
$ \Hom(L, \Z) \semi G $.
If $\hat \rho$~is symmorphic, then
there is some~$\hat \rho_1$, indistinguishable from~$\hat \rho$, such that
the phase function of~$\hat \rho_1$ is zero.  Then \eqref{eqn:symmetry}~shows
that
$ \hat \rho_1 \comp g = \hat \rho_1 $
for all
$ g \in G $.

\begin{Def} \label{def:invariant}
Let $L$~be a quasilattice in~$\R^{d*}$ and let $G$~be a subgroup of the
holohedry group~$G_L$.
A \textbf{gauge invariant} of the pair~$(G, L)$ is a function
$ f \from H^1(G,\hat L) \to \C $.
If $G$~is finite, then a \textbf{fundamental gauge invariant} is a
homomorphism
$ f \from H^1(G,\hat L) \to \C\cross $.
\end{Def}

Thus a gauge invariant assigns a number to each phase function,
or to each Fourier quasicrystal~$\hat \rho$ on~$L$ whose point group
contains~$G$,
and that number depends only on the gauge-equivalence class.
The set of all gauge invariants forms a vector space.
Suppose that $G$~is a finite group.
It follows from Theorem~\ref{thm:dual} that
$H^1(G,\hat L)$ is a finite Abelian group.  Therefore,
this vector space has finite dimension, and the set of characters
$ H^1(G,\hat L) \to \C\cross $
forms a basis.
This explains the term \textit{fundamental gauge invariant}.
Any such character factors through the exponential map
$ e^{2\pi i x} \from \Q/\Z \to \C\cross $,
so we also refer to any homomorphism
$ H^1(G,\hat L) \to \Q/\Z $
as a fundamental gauge invariant.
In these terms, Theorem~\ref{thm:dual} identifies the set of fundamental gauge
invariants as the homology group~$H_1(G,L)$.

\section{Connections with Real-Space Quasicrystals} \label{sec:realqc}

For this section, assume that $\rho$~is a function on~$\R^d$ defined by an
\textit{absolutely convergent} series of the form~\eqref{eqn:formalFS}.
Other authors, such as de Bruijn~\cite{deBruijn81} and Hof~\cite{Hof}, have
considered the general problem of associating Fourier series to quasicrystals,
and work continues on this question.  This paper deals with what to do
\textit{after} obtaining the function~$\hat \rho$ on Fourier space, so the
purpose of this section is to provide a simple analytic setting to illustrate
this theory, not a comprehensive one.
Of course, if the formal series~\eqref{eqn:formalFS} converges in any sense,
then only finitely many terms can have absolute value greater than a given
positive~$\eps$.  Keeping only these terms gives a truncation of the series, or
approximation of~$\rho$, that is certainly absolutely convergent.
Taking~$\eps$ small enough, or taking sufficiently many terms, should give an
approximation that has the same symmetry type as the original.

As in \S\ref{sec:definitions}, the terminology largely follows~\cite{Dr-M}.

\begin{Def} \label{def:density}
A \textbf{density function} is any function
$ \rho \from \R^d \to \C $
given by an absolutely convergent series~\eqref{eqn:formalFS},
where $\hat \rho$~is a Fourier quasicrystal.
\end{Def}

Think of a density function as describing the electron density or mass density
of a physical quasicrystal.  One could also refer to~$\rho$ itself as a
quasicrystal.

Under the hypothesis of absolute convergence, it is easy to see that
the Fourier quasicrystal
$\hat \rho$~can be recovered from
the density function~$\rho$.
Let $C(r)$ denote the cube of side~$r$, centered at the origin,
in~$\R^d$.  Multiplying~\eqref{eqn:formalFS} by~$e^{-2\pi i k' \cdot x}$, the
series is still absolutely and uniformly convergent.  Averaging over~$C(r)$
gives
\begin{equation}
	\label{eqn:Fcoeff}
	\frac{1}{r^d} \int_{C(r)} \rho(x) \cdot e^{-2\pi i k' \cdot x} \,dx
	= \sum_{k \in L} \hat \rho(k) \cdot
	\frac{1}{r^d} \int_{C(r)} e^{2\pi i (k - k') \cdot x} \,dx ,
\end{equation}
which converges absolutely and uniformly in~$r$.  Taking the limit as
$ r \to \infty $
gives~$\hat \rho(k') $.

Define the positionally-averaged $n^{\mbox{\footnotesize th}}$-order
\textbf{autocorrelation function} of the density function~$\rho$ to be
\begin{equation} \label{eqn:autocorrelation}
	\rho^{(n)}(x_1, \ldots, x_n)
	= \frac{1}{r^d} \int_{C(r)} 
	\rho(x_1 - x) \cdots \rho(x_n - x) \,dx .
\end{equation}
Since the product of absolutely convergent series is absolutely convergent,
$ \rho(x_1 - x) \cdots \rho(x_n - x) $
is represented by an absolutely convergent series of the
form~\eqref{eqn:formalFS}, and the argument of the preceding paragraph shows
that the same is true of the autocorrelation function:
\begin{equation} \label{eqn:autocorrelationFS}
	\rho^{(n)}(x_1, \ldots, x_n)
	= \sum_{\substack{k_1, \ldots, k_n \in L \\
		k_1 + \cdots + k_n = 0}}
		\hat \rho(k_1) \cdots \hat \rho(k_n)
		e^{2\pi i (k_1 \cdot x_1 + \cdots + k_n \cdot x_n)} .
\end{equation}

Two density functions
$\rho_1$,~$\rho_2 \from \R^d \to \C$
are called \textbf{indistinguishable} if their autocorrelation functions are
the same.
Mermin~\cite{Mermin92b} and others have
argued that using this criterion, rather than
considering identity of density functions, is the most important theoretical
difference between the Fourier-space approach and traditional crystallography.
If two Fourier quasicrystals
$ \hat \rho_1 $~and~$ \hat \rho_2 $
are indistinguishable (as defined in \S\ref{sec:definitions}), then the
corresponding density functions
$ \rho_1 $~and~$ \rho_2 $
are as well.
It follows that a symmetry of~$\hat \rho$ is a rotation~$g$ such that
$\rho \comp g$~is indistinguishable from~$\rho$, or a macroscopic symmetry
of~$\rho$.
Define the point group of the density function~$\rho$ to be the same as the
point group of the corresponding Fourier quasicrystal~$\hat \rho$.

If the density function~$\rho$ describes a periodic crystal, then $\rho$
is periodic with respect to a lattice
$ \T \cont \R^d $,
and $L$ is dual to~$\T$ (assuming that $\T$~is the lattice of all periods
of~$\rho$).
In this case, $L$~is a lattice, so a gauge function
(an element of
$ \hat L = \Hom(L, \R/\Z) = \R^d / \T $)
is determined by a translation on~$\R^d$, and a symmetry of~$L$ is an
orthogonal transformation of~$\R^d$ that takes~$\T$ to
itself.  In other words, the holohedry group~$G_L$ is the quotient of the
\textbf{space group}~$\mathcal G_{\T}$ of~$\T$---the group of isometries that
preserve~$\T$---by the subgroup of translations corresponding to
elements of~$\T$.
\relax From this point of view, the action of~$G_L$ on~$\T$ is induced by the
conjugation action of~$\mathcal G_{\T}$ on its subgroup of translations.
The density function~$\rho$ can be thought of as additional structure, or
``decoration,'' on the lattice~$\T$.
The space group of~$\rho$
is the group~$\mathcal{G}$
of all isometries~$\g$ that respect this additional structure,
$ \rho \comp \g = \rho $,
and the point group
$ G = \mathcal{G} / \T $
is a subgroup of~$G_L$.

Still in the periodic case,
$ \mathcal G \cong \T \times G $
as a set.
The group structure of~$\mathcal G$ can be recovered from the conjugation
action of~$G$ on~$\T$ and an element of~$H^2(G,\T)$
(\cite{Hiller}, \cite{AJ-I}, \cite[\S IV.3]{Brown}, or \cite[\S2]{C-F}).
As Hiller points out in~\cite{Hiller},
the boundary map of the long exact sequence associated to
$ 0 \to \T \to \R^d \to \R^d/\T \to 0 $
gives an isomorphism of
$ H^2(G,\T) $
with
$ H^1(G, \R^d/\T) $.
Since
\( \R^d / \T \cong \Hom(L, \R/\Z) \),
this is the cohomology group considered in
Definition~\ref{def:symmetrytype}.

In the general case, turn these definitions around as in~\cite{Dr-M}.  Start
with the quasilattice
$ L \cont \R^{d*} $
and define
$ \T = \Hom(L, \Z) $,
naturally embedded in
$ V = \Hom(L, \R) $.
In the aperiodic case, $\dim V = \rank L > d $.
This gives a coordinate-free description of the
\textbf{superspace}~$V$.  In this context, the group~$\mathcal G$
defined by the cohomology class
$ \{ \Phi \} \in H^1(G, \hat L) \cong H^2(G, \T) $
is often called a superspace group, but
this paper uses the term \textit{space group}.

It is not needed in this paper, but one often considers~$\mathcal G$ as a
crystallographic group of isometries of~$V$.  In order to do so, one must
define a Euclidean inner product on~$V$, a point neglected in~\cite{Dr-M}.
Since $L$~spans~$\R^{d*}$, the inclusion
$ L \cont \R^{d*} $
leads to a natural inclusion
$ \R^d \cont V $,
compatible with the action of~$G$.  Take any positive-definite inner product
on~$V$ that extends the standard one on~$\R^d$, and average over~$G$.  This
gives a $G$-invariant inner product on~$V$ that restricts to the usual one
on~$\R^d$, as required.  Not all choices disappear during the averaging
process: if one views the action of~$G$ on~$V$ as a group representation,
each irreducible
subrepresentation (outside of~$\R^d$) can be given an independent scale
factor, and isomorphic irreducible subrepresentations may or may not be
orthogonal.
Any such inner product on~$V$ leads to an orthogonal projection
$ V \to \R^d $,
and the image of~$\T$ under this projection will be a quasilattice.

Two other models of aperiodic quasicrystals start with a lattice
$ \mathcal T \cont V \cong \R^D $
and an embedding
$ \R^d \into V $
that meets~$\mathcal T$ in at most one point.
Taking
$ L = \Hom(T, \Z) $,
one can think of these data in the terms described above.
The ``cut-and-project'' model takes a particular ``slice''
$ S \subset \mathcal T $
and a projection
$ p \from V \to \R^d $;
the set~$p(S)$ is considered a quasicrystal.
The other model takes a tiling
of~$V$, periodic with respect to~$\mathcal T$, and intersects this tiling
with~$\R^d$.
In this variant, the set of vertices of the resulting tiling of~$\R^d$ is the
model of a quasicrystal.  In either case, a suitably general theory of the
Fourier transform (see \cite{deBruijn86a}~or~\cite{Hof-CMP})
applied to the sum of delta functions at points of the quasicrystal leads to
a set of Fourier coefficients~$\hat \rho(k)$ for
$ k \in L $.
The series
$ \sum_{k} |\hat \rho(k)| $
need not converge, so the results of this section do not apply, but the
hypothesis of absolute convergence is not used in the rest of this paper.

\section{Classification} \label{sec:classify}

The terminology in this section mostly follows \cite{Dr-M}.  In
\S\ref{sec:definitions}, a symmetry type was defined to be a
triple
$ (G, L, \{\Phi\}) $,
where $G$~is a finite subgroup of the orthogonal group~$\Orth(d)$, $L$~is a
quasilattice in~$\R^{d*}$ symmetric under~$G$, and $\{\Phi\}$~is a cohomology
class in
$ H^1(G, \hat L) $.
A symmetry type corresponds to a space group, although the (algebraic
structure of the) space group determines only the algebraic structure of the
quasilattice, not its embedding in~$\R^{d*}$.
This section defines when two symmetry types should be considered equivalent
and describes a programme for classifying them.
Since equivalent symmetry types have isomorphic space groups, we usually talk
of classifying space groups.

\begin{Def} \label{def:acc}
Two pairs $(G_1,L_1)$ and~$(G_2,L_2)$ are in
the same \textbf{arithmetic crystal class} if there are a proper rotation
$ r \in \SO(d) $
and an isomorphism
$ f \from L_1 \to L_2 $
as Abelian groups such that
$ G_2 = r G_1 r^{-1} $
and
$ f(kg) = f(k) \cdot r g r^{-1} $
for all
$ k \in L_1 $
and
$ g \in G_1 $.
\end{Def}

Consider first the case
$ f(k) = k r^{-1} $.
Requiring
$ r \in \SO(d) $
means that, in the case $d=2$, mirror-image quasilattices are not necessarily
in the same arithmetic crystal class~\cite{MRW46}.
Next suppose that $r$~is the identity.
Since $f$~is not required to extend to a continuous map on~$\R^{d*}$, this
allows for continuous families of quasilattices all in the same arithmetic
crystal class (see Note~8 in \cite{Dr-M}).

\begin{Def} \label{def:sgt}
Two symmetry types
$ (G_1, L_1, \{\Phi_1\}) $
and
$ (G_2, L_2, \{\Phi_2\}) $
are in the same \textbf{space-group type} if $(G_1,L_1)$ and~$(G_2,L_2)$
are in the same arithmetic crystal class and it is possible to choose
$r$~and~$f$ as in Definition~\ref{def:acc} in such a way that
$ \{ \Phi_2 \} = \{ f \comp \Phi_1 \comp c_r \} \in H^1(G_2, \widehat{L_2})
$,
where
$ c_r \from G_2 \to G_1 $
is the conjugation map
$ c_r(g) = r^{-1} g r $.
\end{Def}

One possibility, which does not occur with discrete lattices, is that
$ G_1 = G_2 $,
$ L_1 = L_2 $,
and $f$~is a non-trivial dilation.  For example, identifying the complex plane
with~$\R^{2*}$, let
$ \zeta = e^{2\pi i/5} $
and
$ L = \Z[\zeta] $.
Then
$ f(x) = (\zeta + \zeta^{-1})x $
gives an isomorphism of~$L$ onto itself, and
$ \zeta + \zeta^{-1} = (\sqrt{5}-1)/2 $
is a real number between 0~and~1.
The identification of the symmetry types described by $\Phi$
and~$f \comp \Phi$ is sometimes called
\textbf{scale invariance}~\cite{RWM88b}.

One of the main goals of crystallography is to classify the possible
space-group types.  This paper considers only the case where $G$~is finite;
see~\cite{Piun-survey} for examples and further discussion of quasilattices
with infinite holohedry groups.  We propose the following classification
programme:
\begin{enumerate}
\item \label{classify:group}
Find all finite groups
$ G \cont \Orth(d) $,
up to conjugation by~$\SO(d)$.
\item \label{classify:lattice}
For each point group~$G$, classify the quasilattices~$L$ that are stable
under~$G$.
\item \label{classify:phase}
Calculate the cohomology group
$ H^1(G,\hat L) $.
\item \label{classify:equiv}
Consider the action of automorphisms of the pair $(G,L)$ on this
cohomology group.  That is, consider $f$~and~$r$ as above in the case
$ G_1 = G_2 = G $
and
$ L_1 = L_2 = L $.
\end{enumerate}

The first step is well known in dimensions two and three.
If $d=2$, such a group is either cyclic or dihedral; in the latter case, one
can take the $x$-axis as one of the mirror lines.
If $d=3$, see, for example,
\cite[Appendices A~and~B]{Weyl}.
In two dimensions, Step~2 can be done using ideas from integer representation
theory, especially the theory of twisted group algebras:  see
\cite[\S28]{C-R}~and~\cite{Theohari}.
The authors are working on a paper that explains these ideas in simpler terms.
The results of the current paper are useful for the third step.
Sections \ref{sec:cyclic}~and~\ref{sec:res-inf} compute
$ H^1(G,\hat L) $
in the case $d=2$.
The authors hope to study the case $d=3$ in future work.
The final step of the classification is actually quite controversial; perhaps
it is safest to say that, for some applications, it is appropriate to identify
the cohomology class of~$\Phi$ with that
of~$f \comp \Phi \comp c_r$.
In any event, this step will depend on the solution of
Step~\ref{classify:lattice}.

We summarize this approach to classification with the phrase, ``$G$~first,
then~$L$.''  We feel this is appropriate in the Fourier-space approach to
quasicrystals, since the symmetry of an X-ray diffraction pattern is more
apparent than the rank of the quasilattice.  (The diffraction pattern may have
more symmetries than the point group.)  Perhaps more significantly,
quasilattices are much more varied than discrete lattices, so it is helpful
to impose some order by first specifying the point group, as in the first step
of the classification programme.  For these reasons, Definition~\ref{def:acc}
differs from the definition in~\cite{Dr-M}: Dr\"ager and Mermin say that two
quasicrystals are in the same arithmetic crystal class only if the holohedry
groups
$ G_{L_1} $~and~$ G_{L_2} $
as well as the point groups
$ G_1 $~and~$ G_2 $
are related by the proper rotation~$r$ in Definition~\ref{def:acc} (although
Note~8 in~\cite{Dr-M} partially contradicts this).
In a sense, the ``$G$~first'' approach is not really new:  several papers,
such as~\cite{RWM}, assume that the quasilattice~$L$ has minimal rank
consistent with its rotational symmetry, which is very natural from this point
of view.

Sometimes the ``$G$~first'' approach requires very minor adjustments.  For
example, \cite{RWM}~discusses the two-dimensional lattice~$L$ of equilateral
triangles, symmetric under a six-fold rotation.  Fixing the lattice, there are
two distinct copies of the dihedral group~$D_3$ (or~$3m$ in International
crystallographic notation) inside the holohedry group~$G_L$:  one contains
mirror lines through the shortest vectors, and the other contains mirror lines
between the shortest vectors.  In the ``$G$~first'' approach, one fixes the
dihedral group~$D_3$ containing the reflection in the $x$-axis.  There are
then two types of lattice, one with its shortest vectors along the mirror
lines and one with its shortest vectors between the mirror lines.
Evidently, these are two different ways of describing the same situation.

A more significant difference between the two approaches emerges when
considering the square lattice.  Here, if the lattice is fixed, then there is
only one dihedral group~$D_4$, with mirror lines both through and between the
shortest vectors.  However, if the group is fixed, and one of the mirror lines
is identified with the $x$-axis, then there are two square lattices to
consider:  one with a shortest vector along the $x$-axis and one with shortest
vector along the $45^\circ$~line.
In our classification programme, these two lattices are considered distinct
until the final step.
This distinction is essential when classifying quasilattices of non-minimal
rank, as discussed in~\cite{fisherrabsoninprep}.
Similar remarks apply when considering~$D_N$ where $N$~is any higher power
of~$2$.

\section{Higher Cohomology is Torsion} \label{sec:torsion}

In this section, assume that $G$~is a finite group acting on the right on a
free Abelian group~$L$ of finite rank.  Let
\begin{equation}
   N = \#G.
\end{equation}
A standard theorem (\cite[\S VI.5]{Brown} or~\cite[\S6, Cor.~2]{C-F})
states that the Tate cohomology groups~$\hat H^i(G, M)$ are torsion, killed
by~$N$.  In particular, the homology group
$ H_1(G,M) = \hat H^{-2}(G,M) $
and the cohomology group
$ H^1(G,M) = \hat H^1(G,M) $
are killed
by~$N$.  In the crystallographic literature so far, the following consequence
has been noted only in the cases where the cohomology group has been
explicitly calculated \cite{RWM, Rabson91, Mermin92a}.
Give the Pontrjagin dual
$ \hat L = \Hom(L, \R/\Z) $
the standard left $G$-action,
$ (g \chi)(k) = \chi(k g) $.

\begin{prop} \label{prop:torsion}
  Given a cohomology class in $H^1(G,\hat L)$,
  one can choose a representative cocycle\/~$\Phi$ so that
  $$ \Phi_g(k) \in (\tfrac1N \Z) / \Z	\qquad (g\in G,\ k\in L) . $$
  That is, with a suitable choice of gauge, any phase function takes values
  in\/
  $ (\tfrac{1}{N}\Z) / \Z $.
\end{prop}

\begin{proof}
Since the cohomology class of~$\Phi$ is killed by~$N$,  $N\Phi$~is a
coboundary.  In other words, there is a
$ \chi \in \hat L $
such that
$ N \Phi = d \chi $, where $d$ is the coboundary operator.
Since
$ \hat L \cong (\R/\Z)^{\rank(L)} $,
one can choose
$ \chi_1 \in \hat L $
such that
$ N \chi_1 = \chi $.
Let
$ \Phi^{(1)} = \Phi - d \chi_1 $.
Then  $\Phi^{(1)}$  is in the same cohomology class as~$\Phi$, and
$ N \Phi^{(1)} = N \Phi - d \chi = 0 $.
In terms of  $g\in G$  and  $k\in L$,  this means that
$ N \Phi^{(1)}_g(k) = 0 \in \R/\Z $,
or
$ \Phi^{(1)}_g(k) \in (\tfrac1N \Z) / \Z $.
\end{proof}

\begin{notn} \label{notn:Qdual}
If  $A$  is any Abelian group, denote the dual of~$A$ by
$$ A' = \Hom(A, \Q/\Z) . $$
If $A$~is a right $G$-module, then give~$A'$ the standard left $G$-module
structure:
for $g\in G$, $\phi\in A'$, and any $a\in A$,
$g\phi$  is defined by
$ (g\phi)(a) = \phi(a g) $.
If $A$ is a left $G$-module, then $g\phi$  is defined by
$ (g\phi)(a) = \phi(g^{-1} a) $,
or
$ (g\phi)(g a) = \phi(a) $.
\end{notn}

\begin{prop} \label{prop:isom}
There is a natural isomorphism
$ H^1(G, L') \isom H^1(G, \hat L) $.
\end{prop}

\begin{proof}
Since  $L$  is a finitely generated free Abelian group, the short exact
sequence
$ 0 \to \Q/\Z \to \R/\Z \to \R/\Q \to 0 $
leads to the short exact sequence
\begin{equation}
   0 \to L' \to \hat L \to \Hom(L,\R/\Q) \to 0 ,
\end{equation}
and
$ \Hom(L,\R/\Q) \cong (\R/\Q)^r $,
with
$ r = \rank(L) $.
Since  $\R/\Q$  is uniquely divisible, its Tate cohomology groups vanish, so
the long exact sequence of Tate cohomology gives
\begin{equation}
   0 = \hat H^0(G, \Hom(L,\R/\Q)) \to H^1(G, L') \to H^1(G, \hat L) \to 0 .
  \hfill \qed
\end{equation}
\renewcommand{\qed}{}
\end{proof}

We need this proposition to apply the duality theorem we quote in
\S\ref{sec:dual}, which
is stated in terms of~$L'$.  Note that surjectivity in Prop~\ref{prop:isom},
but not injectivity, also follows from Proposition~\ref{prop:torsion}.

\begin{remark}
If
$\rho_1$~and~$\rho_2$
are indistinguishable Fourier quasicrystals, then Definition~\ref{def:gauge}
requires
$ \hat\rho_2(k) = e^{2\pi i \chi(k)} \hat \rho_1(k) $,
where
$ \chi \in \hat L $.
This implies that
$ \chi(k) \in \R/\Z $,
so that
$ | \hat \rho_1(k) | = | \hat \rho_2(k) | $.
If one were to relax this condition, one would take
$ \chi \from L \to \C/\Z $,
so that
$ e^{2\pi i \chi(k) } $
could be any non-zero complex number.  Making the corresponding change in
Definition~\ref{def:symmetry},
one would consider cohomology with coefficients in
$ \Hom(L,\C/\Z) $
instead of~$\hat L$.
The analogues of Propositions \ref{prop:torsion}~and~\ref{prop:isom} would
still hold, so
\( H^1 \bigl( G, \Hom(L, \C/\Z) \bigr) \cong H^1(G, \hat L) \).
In other words, the alternative definition of indistinguishability does not
lead to any new symmetry types, and a Fourier quasicrystal~$\hat \rho_1$ with
point group~$G$ under the alternative definition is indistinguishable (in the
alternative sense) from a Fourier quasicrystal~$\hat \rho_2$ that has point
group~$G$ using either definition.
\end{remark}

\section{Cohomology is Dual to Homology} \label{sec:dual}

In this section, assume that $G$~is a finite group acting on the right on a
finitely generated Abelian group~$L$.  In particular, this implies that
$H_1(G,L)$ is finite.

In \S\ref{sec:definitions}, we observed that any gauge invariant
$ f \from H^1(G, \hat L) \to \C $
can be expressed in terms of the fundamental gauge invariants, the
homomorphisms
$ H^1(G, \hat L) \to \Q/\Z $.
According to Proposition~\ref{prop:isom}, the set of fundamental gauge
invariants is
$ H^1(G, L')' $
(\cf~Notation~\ref{notn:Qdual}).
We now interpret this set as a homology group.  If  $M$~is a right $G$-module
then write $1$-chains, or elements of
$ M \tensor \Z G $,
as
$ c = \sum_g m_g [g] $,
where $g \in G$ and $m_g \in M$.
The boundary map is defined by
\begin{equation} \label{eqn:1-boundary}
   \del(m [g]) = m g - m .
\end{equation}
For details, see \cite[\S III.1]{Brown} or \cite[\S3]{C-F}.

\begin{thm} \label{thm:dual}
Let $G$~be a finite group, and let $L$~be a finitely generated Abelian group
on which $G$~acts.  Let $\hat L$,
$ H_1(G, L)^{\wedge} $,
and
$ H^1(G, \hat L)^{\wedge} $
denote the Pontrjagin duals as in~\eqref{eqn:Pontrjagin}.
There are natural isomorphisms
$ H_1(G, L) \isom H^1(G, \hat L)^{\wedge} $
and
$ H^1(G, \hat L) \isom H_1(G, L)^{\wedge} $,
induced by the duality pairing
\begin{align*}
  H^1(G, \hat L) \times H_1(G, L) &\to \R/\Z \\
  \bigl( \{\Phi\}, \{c\} \bigr) &\mapsto \bkt{\Phi, c}
  	= \sum_{g\in G} \Phi_g(k_g) ,
\end{align*}
where
$ c = \sum_g k_g [g] $.
\end{thm}

\begin{proof}
According to Proposition~\ref{prop:isom}, the natural map from
$ H^1(G, L') $
to
$ H^1(G, \hat L) $
is an isomorphism, where
$ L' = \Hom(L, \Q/\Z) $
as in Notation~\ref{notn:Qdual}.
The finiteness hypotheses on $G$~and~$L$ imply that
$ H_1(G, L) $
is a finite group.  It follows that the pairing in the theorem takes values
in~$\Q/\Z$ and that
$ H_1(G, L)^{\wedge} = H_1(G, L)' $.
Roughly speaking, the finiteness hypotheses imply that one can replace~$\R/\Z$
with~$\Q/\Z$ throughout.
	
According to~\cite[Prop.~VI.7.1]{Brown}, there is a duality pairing between
$ H^1(G, L') $
and
$ H_1(G, L) $
that identifies each with the dual of the other
(in the sense of Notation~\ref{notn:Qdual}).
Up to a sign, this pairing agrees with the one in the statement of the theorem
by
\cite[\S V.3]{Brown} and \cite[\S III.1, Example~3]{Brown}.
In particular, this shows that
$ H^1(G, L') $
is a finite group, so
$ H^1(G, \hat L)^{\wedge}
= H^1(G, L')^{\wedge}
	= H^1(G, L')'
$.
Thus the duality of the theorem is just a restatement of the duality between
$ H^1(G, L') $
and
$ H_1(G, L) $.
\end{proof}

\begin{cor} \label{cor:inv}
The gauge-equivalence class of the phase function~$\Phi$ is determined by the
gauge-invariant rational numbers\/~$\bkt{\Phi,c}$ for
$ c \in H_1(G,L) $.
\end{cor}

\begin{proof}
This is simply a restatement of the injectivity of the map
$ H^1(G, \hat L) \to H_1(G, L)^{\wedge} $.
\end{proof}

\begin{remark} \label{rem:super}
As noted in Section~\ref{sec:realqc}, the view from superspace is that
the class~$\{\Phi\}$ in
$ H^1(G, \hat L) \cong H^2(G, \T) $
describes the space group~$\mathcal G$, an extension of~$G$ by
$ \T = \Hom(L, \Z) $.
Recall that
$ \hat L \cong V/\mathcal T$,
where $V$~denotes the superspace
$ V = \mathcal T \tensor \R $.
As described in~\cite{Hiller},
$ \Phi_g \in V / \T $
is the coset of~$\T$ consisting
of all translations that can be combined with~$g$ to give an element of the
space group~$\mathcal G$.
Theorem~\ref{thm:dual} still applies, so
$ H^1(G, V/\T) $
is dual to
$ H_1(G, L) = H_1 \bigl (G, \Hom(\T, \Z) \bigr) $.

Let us make this duality pairing explicit.  Let
$ c = \sum_g k_g [g] $
be a cycle, with coefficients
$ k_g \in \Hom(\T, \Z) $,
and let $\Phi$~be a cocycle as above.
Choose a basis
$t_1$,~\dots,~$t_n$
of~$\mathcal T$ over~$\Z$; it is also an $\R$-basis of~$V$.  If
$ v = v_1 t_1 + \dots + v_n t_n \in V $
and
$ k \in \Hom(\T, \Z) $, define
$ \bkt{v,k} = v_1 \cdot k(t_1) + \dots + v_n k(t_n) \in \R $.
Similarly, define
$ \bkt{\bar v, k} \in \R/\Z $
if
$ \bar v \in V/\mathcal T $.
Then the duality pairing is defined by
$ \bkt{\Phi, c} = \sum_g \bkt{\Phi_g, k_g} $.
\end{remark}

\begin{remark}
The simplest example of a $1$-chain is
$ c = k [g] $,
with  $k\in L$  and  $g\in G$.  By \eqref{eqn:1-boundary},
this chain is a cycle if and only if  $kg=k$, and in this case the
corresponding gauge invariant is simply~$\Phi_g(k)$.
However, the homology group $H_1(G,L)$ is not always generated by cycles of
this form.
In other words, it is possible for two gauge-inequivalent cocycles
$\Phi^{(1)}$~and~$\Phi^{(2)}$ to have the same ``obvious'' gauge invariants:
\( \Phi^{(1)}_g(k) = \Phi^{(2)}_g(k) \)
whenever $kg=k$.
In fact, of the $230$ classical space groups,
there are two non-symmorphic ones,
denoted $I2_12_12_1$ and
$I2_13$ in international crystallographic notation,
for which all cycles of the form
$ c = k [g] $
are boundaries~\cite{Mermin92a,K-M:1997}.
Since these space groups are non-symmorphic, Theorem~\ref{thm:dual}  shows
that
$H_1(G,L) \ne 0$,
so there must be other cycles.
What is the next simplest cycle one can construct?
Since  $H_1(G,L)$  is killed by  $N=\#G$, any cycle becomes trivial
in~$H_1(G, \tfrac1N L)$,  so it is natural to consider the boundary of a
$2$-chain with values in~$\tfrac1N L$:
if the result happens to have coefficients in~$L$, it is a $1$-cycle
in $H_1(G, L)$.
In the notation of
\cite[\S III.1, Example~3]{Brown}, the
boundary of the $2$-chain  $q [g|h]$
(where $q \in \tfrac1N L$ and $g$,~$h \in G$)
is given by
\begin{equation} \label{eqn:2-boundary}
   \del( q [g|h] ) = (q g) [h] - q [gh] + q [g] .
\end{equation}
This cannot give a non-trivial homology class in~$H_1(G,L)$ by itself.
Perhaps the simplest combination that can is
\begin{equation} \label{eqn:cocycle}
   \del( q [g|h] - q [h|g] )
	= (q g - q) [h] - q ( [gh] - [hg] ) + ( q - q h ) [g] ,
\end{equation}
which will have values in~$L \tensor \Z G$ provided that
$ q g - q $,
$ q h - q \in L $
and
$ gh = hg $.
It is a simple exercise
to calculate the homology groups corresponding to the
two exceptional space groups $I2_12_12_1$ and $I2_13$.
(See~\cite{RabsonFisher02} for one of the two cases.)
In both cases, the homology group is cyclic of order~$2$, generated by the
class of such a cycle.
\end{remark}

Gauge invariants are considered again in~\S\ref{sec:products}.
We conclude this section with a new proof of the result in the appendix
of~\cite{K-M:1999}.
This states that if the ``obvious'' gauge invariants of~$\Phi$ corresponding
to a single $g \in G$ all vanish, then (up to gauge equivalence) $\Phi_g$~is
trivial.  The examples cited above show that one cannot necessarily find a
gauge in which $\Phi_g=0$ simultaneously for all $g\in G$, even if all these
gauge invariants vanish.

\begin{prop} \label{prop:inv}
Let $g\in G$, and let\/
$ \{\Phi\} \in H^1(G,\hat L) $.
If one choice of\/ $\Phi$ satisfies\/  $\Phi_g = 0$  on
$ L^g = \{ k \in L \mid kg = k \} $,
then one can choose\/~$\Phi$ such that
$\Phi_g(k) = 0$  for all $k \in L$.
\end{prop}

\begin{proof}
Let
$ \bkt{g} = \{1, g, \ldots, g^{N-1} \} $
denote the subgroup of~$G$  generated by~$g$.  We claim that  $\Phi$~is
trivial in~$H^1(\bkt{g}, \hat L)$.  By Corollary~\ref{cor:inv}, it suffices to
show that
$ \bkt{ \Phi , c } = 0 $
for all
$ c \in H_1(\bkt{g},L) $.
According to \eqref{eqn:ho-cyclic} below,
$ H_1(\bkt{g},L) = L^g / N_g L $,
where
$ N_g = 1 + g + \cdots + g^{N-1} $.
Therefore the hypothesis  $\Phi_g(L^g)=0$  justifies the claim.

Since $\Phi$~is trivial in~$H^1(\bkt{g},\hat L)$,  there is some
$ \chi \in \hat L $
for which
$ \Phi_g = (d \chi)_g = g \chi - \chi $.
Then
$ \Phi^{(1)} = \Phi - d \chi $
represents the same class in~$H^1(G,\hat L)$, and $\Phi^{(1)}_g=0$.
\end{proof}

\section{Homology and Cohomology of Cyclic Groups} \label{sec:cyclic}

This section and the following one classify the space groups
corresponding to the finite point group~$G$ and the quasilattice~$L$
in two dimensions.
This section discusses cyclic groups, and the next deals with dihedral groups.
The classification applies to
``non-standard''~\cite{MRW46} as well as to ``standard'' quasilattices and
applies whether or not the rank of~$L$ is minimal
given that $L$~is symmetric under~$G$.
Work in progress~\cite{fisherrabsoninprep} classifies the quasilattices
of non-minimal rank (sometimes called \textit{modulated} quasilattices)
symmetric under~$G$.

Let  $G$  be a finite cyclic group, say
\begin{equation}
  G = \bkt{r} = \{ 1, r, \ldots, r^{N-1} \} .
\end{equation}
(If $G$~is a subgroup of $\Orth(2)$~or~$\Orth(3)$, then
the generating element~$r$ might be a rotation or, for $N=2$, a mirror.)
If   $M$  is any left  $G$-module, then think of  $r-1$  and the norm element
\begin{equation}
   N_r = 1 + r + \cdots + r^{N-1}
\end{equation}
in terms of their actions on~$M$:
$ (r-1)x = rx - x $
and
$ N_r x = x + rx + \cdots + r^{N-1} x $.
According to \cite[\S III.1]{Brown} or \cite[\S8]{C-F},
the Tate cohomology groups can, in this case, be computed as the cohomology of
the complex
\begin{equation}
   \cdots \xrightarrow{N_r} M \xrightarrow{r-1} M
	\xrightarrow{N_r} M \xrightarrow{r-1} M
	\xrightarrow{N_r} \cdots \, .
\end{equation}
In particular,
\begin{align}
  H^1(G,M) &= \hat H^1(G,M) = \ker(N_r) / (r-1) M ; \label{eqn:co-cyclic} \\
  H_1(G,M) &= \hat H^{-2}(G,M) = \ker(r-1) / N_r M . \label{eqn:ho-cyclic}
\end{align}
Note that the kernel of~$r-1$ is
$ M^r = \{ x \in M \mid rx = x \} $.

In traditional crystallography, this description of the (co)homology groups
is of limited interest since a two- or three-dimensional rotation that
stabilizes a discrete lattice can
only have order
$1$,~$2$, $3$, $4$, or~$6$.
Quasicrystals can be symmetric under rotations of any order, so these
results become much more useful.

The following proposition shows that
if a two-dimensional point group is
cyclic (of order $N>1$) then the only
corresponding space group is the symmorphic one.
Note that this analysis applies uniformly to any two-dimensional quasilattice.
The case where
$ L \cong \Z[e^{2\pi i/N}] $
is treated in~\cite{RWM}.

\begin{prop} \label{prop:cyclic}
Let  $L \cont \R^{2*}$  be a quasilattice invariant under  $G = \bkt{r}$,  where
$r$~is a rotation of order~$N>1$.  Then  $H^1(G,\hat L)=0$.
\end{prop}

\begin{proof}
It is easier to work with homology of~$L$ than the cohomology of~$\hat L$,
so consider
$H_1(G,L)$.  The only vector in~$\R^{2*}$ fixed by a non-trivial rotation is the
zero vector.  According to \eqref{eqn:ho-cyclic}, this
shows that  $H_1(G,L)=0$.  The result now follows from Theorem~\ref{thm:dual}.
\end{proof}

\section{The Restriction-Inflation Sequence} \label{sec:res-inf}

For this section, let $G$~be a finite group, let
$ H \normal G $
be a normal subgroup, and let
\begin{equation}
 Q = G / H
\end{equation}
denote the quotient.
For any left $G$-module~$M$, the inflation map
$ H^1(Q, M^H) \to H^1(G,M) $
and the restriction map
$ H^1(G,M) \to H^1(H,M) $
fit together to give an exact sequence \cite[\S5]{C-F}
\begin{equation} \label{eqn:res-inf}
  0 \to H^1(Q, M^H) \to H^1(G,M) \to H^1(H,M) .
\end{equation}
This can be viewed as a consequence of the Hochschild-Serre spectral sequence,
as can its homological version \cite[Theorem VII.6.3]{Brown}:
\begin{equation} \label{eqn:homology-SES}
  H_1(H, M) \to H_1(G,M) \to H_1(Q, M_H) \to 0 ,
\end{equation}
where  $M$~is now a right $G$-module and
$M_H$  denotes the quotient of~$M$  by the $H$-submodule generated by
$ \{ xh - x \mid x\in M,\ h \in H \} $.

Let $G \cont \Orth(3)$ be a finite group.  By the classification of such
groups~\cite[Appendices A~and~B]{Weyl},
all but finitely many such~$G$
contain a normal, cyclic subgroup~$H$, generated by a
rotation or a roto-inversion, for which the quotient group $Q=G/H$ has order
$1$,~$2$, or~$4$.  Since $H$~is cyclic, the homology group
$ H_1(H, L) $
can be computed using \eqref{eqn:ho-cyclic}.
In the simplest case, $H$~is generated by a roto-inversion, so
$ H_1(H, L) = 0 $,
and
$ H_1(G, L) = H_1(Q, L_H) $ by~\eqref{eqn:homology-SES}.
We now apply this approach to the two-dimensional case.

\begin{notn}
For the rest of this section, let
$L \cont \R^{2*}$  be a quasilattice invariant under
the dihedral group with $2N$~elements:
\begin{equation}
	G = D_N = \bkt{r,m} \cont \Orth(2) ,
\end{equation}
where $r$~is a rotation of order~$N>1$ and $m$~is a reflection.
Let
\begin{equation}
	H = C_N = \bkt{r}, \qquad D_1 = D_N/H = \{e, \mbar\}
\end{equation}
denote the cyclic subgroup of~$D_N$ and the quotient group.  Let
\begin{equation}
	\zeta = \zeta_N = e^{2\pi i/N} ,
\end{equation}
so that $L$~is a module over the ring of cyclotomic integers~$\Z[\zeta]$, and
note that
$ L_H = L / (1 - \zeta)L $.
Let
\begin{equation} \label{eqn:F2}
	\F_2 = \Z / 2\Z
\end{equation}
denote the field with two elements.
\end{notn}

The results that follow show that $H^1(D_N,\hat L)=0$,
so every space group corresponding to $D_N$~and~$L$ is symmorphic,
unless $N$~is a power of~$2$.
If $N=2^e$, then Theorem~\ref{thm:2D} states that $H^1(D_N,\hat L)$ is a
vector space over
the field with two~elements and counts its dimension.
In other words, still assuming $N = 2^e$, the fundamental invariants all take
the values $0$~and~$1/2$
(modulo~$1$).
In particular, if $L$~has rank~$1$ as a $\Z[\zeta]$-module, then the
cohomology group has exactly two elements:  one corresponds to the symmorphic
space group, and the other corresponds to a non-symmorphic group.
These results were obtained in~\cite{RWM} under the more restrictive
assumption that
$ L \cong \Z[\zeta] $
as a $\Z[\zeta]$-module.

\begin{prop} \label{prop:dihedral}
If $N$~is not a power of\/~$2$, then  $H^1(D_N,\hat L)=0$.
If $N$~is a power of\/~$2$, then\/ $L_H$ is a vector space over\/~$\F_2$.
\end{prop}

\begin{proof}
By Theorem~\ref{thm:dual}, it suffices to compute
$ H_1(D_N, L) $.
By Proposition~\ref{prop:cyclic},
$ H_1(H, L) = 0 $.
Then \eqref{eqn:homology-SES} implies that
$ H_1(D_N, L) \isom H_1(D_1, L_H) $.
According to Lemma~\ref{lemma:unit}, below,  $1 - \zeta$  is a unit unless
$N=p^e$~is a prime
power, in which case its norm is~$p$.
Thus $L_H=0$, and $H_1(D_N,L)=0$,  unless  $N=p^e$.

Suppose now that $N=p^e$.
Then $L_H$ is a vector space over
$ \Z[\zeta] / (1 - \zeta) \cong \F_p $,
the field with $p$~elements.
If $p=2$, this justifies the last claim in the statement.
Now assume that $p$~is odd.
Decompose~$L_H$ into eigenspaces for~$\mbar$:
$ L_H = L_H{}^+ \oplus L_H{}^- $.
On  $L_H{}^+$,
$ N_{\mbar} = 1+\mbar = 2 $,
so
$ N_{\mbar} L_H{}^+ = 2 L_H{}^+ = L_H{}^+ $
(since multiplication by~$2$ is an isomorphism on an $\F_p$-vector space
when $p$ is odd),
and
$ H_1(D_1,L_H{}^+) = 0 $  by~\eqref{eqn:ho-cyclic}.
On the other hand,  $(L_H{}^-)^{\mbar} = 0$,
so
$ H_1(D_1,L_H{}^-) = 0 $
as well.  Therefore,
$ H_1(D_1, L_H) = H_1(D_1, L_H{}^+) \oplus H_1(D_1, L_H{}^-) = 0 $.
\end{proof}

The following lemma is not original, but we do not know a
convenient reference for it.

\begin{lemma} \label{lemma:unit}
Let $N>1$ and let
$ \zeta = \zeta_N = e^{2\pi i/N} $.
If  $N=p^e$  is a prime power, then
$ N_{\Q(\zeta)} (1 - \zeta) = p $;
otherwise,
$ N_{\Q(\zeta)} (1 - \zeta) = 1 $.
Here,
$ N_ {\Q(\zeta)} $
denotes the norm from~$\Q(\zeta)$ to~$\Q$.
\end{lemma}

\begin{proof}
Let  $F_N(x)$  denote the cyclotomic polynomial of order~$N$.  That is,
$F_N(x)$  is the monic, irreducible polynomial whose roots are the primitive
$N$-th roots of unity.%
\footnote{The usual notation for this polynomial is~$\Phi_N(x)$.  In this
paper, $\Phi$~is used to denote a phase function.}
Since these are exactly the conjugates of~$\zeta$ over~$\Q$, it follows that
$$ F_N(1) = \prod_{F_N(\alpha)=0} (1-\alpha)
	= N_ {\Q(\zeta)} (1 - \zeta) .
$$

Since the roots of $x^N-1$  are all the $N$-th roots of unity,
$ x^N - 1 = \prod_{d \mid N} F_d(x) $.
Dividing by $x-1$  and setting   $x=1$  leads to
$ N = \prod_{1 < d \mid N} F_d(1) $.
The lemma now follows by induction on~$N$.
\end{proof}

\begin{notn} \label{notn:Jordan}
Let $M$ be an $n\times n$ matrix over~$\F_2$ such that $M^2=I_n$.  The
Jordan normal form of~$M$ consists of $1\times1$ and $2\times2$ blocks only,
with the number 1 the only possible eigenvalue.  For
example, the Jordan normal form of the standard $2\times2$ reflection
matrix
$\begin{bmatrix}0&1\\ 1&0 \end{bmatrix}$
is
$\begin{bmatrix}1&0\\ 1&1 \end{bmatrix}$.
Let $j_1(M)$ denote the number of $1\times1$ Jordan
blocks and $j_2(M)$ the number of (defective)
$2\times2$ Jordan blocks in the
Jordan normal form of~$M$.  Then
$ j_1(M) + 2 j_2(M) = n $,
and
$ j_1(M) + j_2(M) = n - j_2(M) $
is the dimension of the $1$-eigenspace of~$M$.
\end{notn}

\begin{thm} \label{thm:2D}
Let
$ L \cont \R^{2*} $
be a quasilattice invariant under
$ G = C_N $ or\/~$D_N$.
Then
$ H^1(G, \hat L) = 0 $
unless $G=D_N$ and $N=2^e$, with\/ $e\ge1$.  In this case,
let $M$ be the matrix of~$\mbar$ acting on the\/ $\F_2$-vector space~$L_H$.
Then  $H^1(D_N,\hat L)$  is an\/ $\F_2$-vector space of dimension~$j_1(M)$.
\end{thm}

\begin{proof}
The case $G=C_N$ is considered in Proposition~\ref{prop:cyclic},
so assume that $G=D_N$.
Proposition~\ref{prop:dihedral} shows that the cohomology
group vanishes if $N$~is not a power of~2, so assume now that $N=2^e$.
By Theorem~\ref{thm:dual}, $H^1(D_N, \hat L)$ is dual to $H_1(D_N,L)$,
so it suffices to show that this homology group has the stated form.
By Proposition~\ref{prop:cyclic} and the exact
sequence~\eqref{eqn:homology-SES},
$ H_1(D_N,L) \cong H_1(D_1,L_H) $.

Since $D_1=\{e,\mbar\}$, it follows from \eqref{eqn:ho-cyclic} that
$ H_1(D_1,L_H) \cong (L_H)^\mbar / (1+\mbar)L_H $.
Note that, although $m$~is antilinear, $\mbar$~is linear as a map on~$L_H$, so
Notation~\ref{notn:Jordan} applies.
An easy calculation shows that each $1\times1$ Jordan block contributes a
one-dimensional space to $H_1(D_1,L_H)$ and that each $2\times2$ Jordan block
contributes nothing.
\end{proof}

\begin{remark} \label{rem:Piun}
Piunikhin \cite{Piun-relation} recognized the cohomological interpretation of
phase functions and noted that
$ H^1(G, \hat L) $
describes an extension
$ 1 \to \Hom(L, \Z) \to \mathcal{G} \to G \to 1 $.
Piunikhin implicitly assumes that $\Hom(L, \Z)$ is a quasilattice if $L$~is.  
We do not see a natural way to regard $\Hom(L, \Z)$ as a quasilattice, but as
we
described in Section~\ref{sec:realqc} there are many ways to do so, all in the
same arithmetic crystal class.  Given this, $\mathcal{G}$ is a
quasicrystallographic group in the sense of Novikov', and Piunikhin's
classification~\cite{Piun} of such groups (with finite point group~$G$) in two
dimensions answers the same question as Theorem~\ref{thm:2D} here.
Our proof is different, and the description here of the dimension of
$ H^1(G, \hat L) $
is simpler than Piunikhin's:  he describes the classification in
terms of an anti-linear involution on~$L$, while Theorem~\ref{thm:2D} uses
linear
algebra over~$\F_2$.

This is a good place to point out a misstatement in~\cite{Piun}.
Let $T$ denote a quasilattice in $\R^2$, invariant under $D_N$ with $N$~even.
Then $T$ can be thought of as a $\Z[\zeta]$-module.  Let $I$~denote the
anti-linear involution of~$T$ corresponding to a mirror reflection
$ m \in D_N $.
Piunikhin describes the correspondence between isomorphism classes of such
pairs $(T,I)$ and arithmetic crystal classes of such quasilattices as being
2-to-1, since $(T,I)$ and $(T, \zeta I)$ are not isomorphic as modules with
involution.  This is not true in general:  for example, consider
$ T = \Z[\zeta_{2N}] $
as a $\Z[\zeta]$-module, and let $I$~be complex conjugation.
\end{remark}

\section{Cohomology Products and Physical Implications} \label{sec:products}

This section uses the notation and hypotheses of Sections
\ref{sec:definitions}~and~\ref{sec:realqc}.
For physical applications, work in dimension $d=3$.
In particular, $G$~is a finite subgroup of the orthogonal group~$\Orth(3)$,
$\hat \rho$ is a Fourier quasicrystal on the quasilattice
\( L \cont \R^{3*} \),
and $\Phi$~is the corresponding phase function,
or cocycle, representing a cohomology class in~$H^1(G,\hat L)$.
So far, we have considered only geometric aspects of crystallography.  This
section discusses some physical implications.  We show that the language of
group cohomology, especially the cup and cap products, provides a convenient
framework for making connections between phase functions and group
representations.

Given a map
$ M \tensor_\Z N \to P $
of $G$-modules, one constructs the cup product
\begin{equation}
   H^m(G,M) \times H^n(G,N) \xrightarrow{\cup} H^{m+n}(G,P)
\end{equation}
and, if $m\le n$, the cap product
\begin{equation}
   H^m(G,M) \times H_n(G,N) \xrightarrow{\cap} H_{n-m}(G,P)
\end{equation}
as in \cite[\S V.3]{Brown} or \cite[\S 7]{C-F}.
If $m=n$ and $M=N'$, then (up to a sign) the cap product is
the same as the duality pairing of \S\ref{sec:dual}, with
$ H_0(G,P) = H_0(G, \Q/\Z) = \Q/\Z $.
Among the various properties enjoyed by these two products are two associative
laws:
$ (\alpha \cup \beta) \cup \gamma = \alpha \cup (\beta \cup \gamma) $
and
$ (\alpha \cup \beta) \cap c = \alpha \cap (\beta \cap c) $
if
$ \alpha$,~$\beta$, $\gamma \in H^*$,
$ c \in H_* $,
and the coefficients are chosen compatibly.

Recall from Sections \ref{sec:definitions}~and~\ref{sec:dual} that
elements of
$ H^1(G, \hat L) $
describe symmetry types of quasicrystals and that
$ H_1(G,L) $
is the set of fundamental gauge invariants.
These are related to several other (co)homology groups by the cup and cap
products, and these groups also have important interpretations.

Consider
$ H^1(G, L) $.
If
$ q \in \R^{3*} $
satisfies
$ q g - q \in L $
for all
$ g \in G $,
then
$ \sigma(g) = q g^{-1} - q $
is a cocycle with values in~$L$.  From the long exact sequence
\cite[Prop.~0.4]{Brown} or~\cite[Theorem~1]{C-F} associated to
$ 0 \to L \to \R^{3*} \to \R^{3*}/L \to 0 $,
it follows that any class in
$ H^1(G, L) $
is represented by such a cocycle.

Next, recall the interpretation of the cohomology group
$ H^2(G, \Q/\Z) $.
A \textbf{projective representation}, or
\textbf{ray representation}, of~$G$ is a homomorphism into the
projective linear group~$\text{PGL}(n)$, just as an ordinary
representation is a homomorphism into the general linear
group~$\text{GL}(n)$.  One associates
to each projective representation
a $2$-cocycle, or \textbf{factor system}, with values in~$\C\cross$;
the factor system depends on additional choices, but its cohomology class
in the \textit{Schur multiplier}
$ H^2(G, \C\cross) $
depends only on the representation.
One standard reference is \cite[\S11.E]{C-R}.  In fact, this theory
is one of the main precursors of group cohomology.
Since $G$ is a finite group, the exponential map
$ \Q/\Z \to \C\cross $
gives an isomorphism
$ H^2(G, \Q/\Z) \isom H^2(G, \C\cross) $
(\cf~\S\ref{sec:torsion}).
The same duality theorem~\cite[Prop.~VI.7.1]{Brown} cited in the proof of
Theorem~\ref{thm:dual} shows that
$ H_2(G, \Z) $
is dual to
$ H^2(G, \Q/\Z) $,
so $2$-cycles with integer coefficients can be thought of as invariants of
factor systems.

We are now ready to discuss physical applications.
Let $\Phi\in H^1(G,\hat L)$  be non-trivial, so that it represents a
non-symmorphic space group.  By Theorem~\ref{thm:dual}, there is some
$ c \in H_1(G,L) $
such that
$ \bkt{\Phi,c} \ne 0 $.
It is reasonable to hope that there is a physical way to distinguish a
non-symmorphic quasicrystal from a symmorphic one, so
one expects such a non-trivial gauge invariant to have physical implications.
If $c$ is represented by a cycle of the form $k[g]$ (with $k\in L$, $g\in G$,
and $kg=k$), then this is well known.
If
$ \hat \rho \from L \to \R $
is any function transforming as in~\eqref{eqn:symmetry} and
$ \Phi_g(k) = \bkt{\Phi,k[g]} \ne 0 $,
then
$ \hat \rho(k) = 0 $.
This is observed as a dark spot in the X-ray-diffraction pattern and is called
a \textbf{systematic extinction}.

Not every gauge invariant is of the above form.  Suppose that
$g$,~$h\in G$
and
$ q \in \R^{3*} $
satisfy
\begin{enumerate}
\renewcommand{\labelenumi}{(\roman{enumi})}
    \item $g h = h g $;
    \item $k_g = q g - q $, $k_h = q h - q \in L $;
    \item $\Phi_{g}(k_h) - \Phi_{h}(k_g) \ne 0 $.
\end{enumerate}
Then $k_g[h] - k_h[g]$ represents a non-trivial homology class;
\cf~\eqref{eqn:cocycle}.
In this situation
K\"onig and Mermin \cite{K-M:1997} describe a projective representation of
$ H = \bkt{g,h} \cont G $
that commutes with the Hamiltonian~$h_q$ corresponding to the wave
vector~$q$ and the potential of the crystal.
Therefore, every eigenspace of the Hamiltonian is a projective
subrepresentation, with the same factor system:
$ (g,h) \mapsto \Phi_h(q g - q) $.
K\"onig and Mermin note that the quantity~(iii) is gauge invariant and, since
it does not vanish, this shows that the projective representation (on each
eigenspace) is not equivalent to an ordinary representation.  In particular,
each eigenspace of the Hamiltonian has dimension greater than one, since
one-dimensional projective representations have trivial factor systems.
This is
expressed by saying that each energy level of~$h_q$ is degenerate, and
the phenomenon is sometimes called \textbf{band sticking}.

We interpret part of this argument as follows.  Let
$ H = \bkt{g,h} $.
Then (i) implies that
$ c = [g|h] - [h|g] $
is a $2$-cycle with coefficients in~$\Z$
(\cf~\eqref{eqn:cocycle}
and~\cite[\S~II.3, Exercise~1]{Brown}).
Condition~(ii) implies that
$ \sigma(g) = k_{g^{-1}} = q g^{-1} - q $
represents a class in
$ H^1(H,L) $,
so
$ \sigma \cap c = k_g[h] - k_h[g] $
represents a class in
$ H_1(G, L) $.
Thus (iii)~means that
$ \bkt{ \Phi, \sigma \cap c } \ne 0 $.
Since
$\Phi\cup\sigma$ is the $2$-cocycle
$ (g,h) \mapsto \Phi_h(k_g) $,
the following proposition applies.

\begin{prop} \label{prop:KM}
Let $H$~be a finite subgroup of $\Orth(3)$ and let
$ L \cont \R^{3*} $
be a quasilattice stable under~$H$.  Let
$ c \in H_2(H, \Z) $,
$ \sigma \in H^1(H, L) $,
and\/
$ \Phi \in H^1(H, \hat L) $
be given.  Then
$$ \bkt{\Phi, \sigma \cap c} = \bkt{\Phi \cup \sigma, c} . $$
In particular, if this quantity is non-zero, then\/ $\Phi$ represents a
non-symmorphic space group, and the factor system\/
$ \Phi \cup \sigma $
is non-trivial.
\end{prop}

\begin{proof}
This follows from associativity of cup and cap products, as described above,
and from the compatibility
$ \bkt{\alpha, \beta} = - \alpha \cap \beta $
between the duality pairing and the cup product.
\end{proof}

Computations of
$ H_1(G, L) $
using the methods described in Section~\ref{sec:res-inf} suggest that this
homology group is usually generated by cycles of the form
$ \sigma \cap c $
as described in the proposition and those of the form
$ k[g] $
with
$ kg = k $.
The hypotheses of the proposition are thus less restrictive than they seem at
first glance.
There are, however, examples where
$ H_1(G, L) $
is not generated by such cycles.
It is not clear what, if any, physical consequences there are in such cases.
The authors hope to return to both of these points in future papers.

\section{Acknowledgements}

This work has been supported in part by the National Science Foundation
through grants DMS-0204823 and DMS-0204845.
DAR is a Cottrell Scholar of Research Corporation and
is grateful to the Boston College mathematics department,
whose hospitality in the summer of 2000 made this collaboration possible.
Both authors sincerely thank the referee of a previous version of this paper,
who provided several important
references as well as making many suggestions for improvement.

\bibliographystyle{iop}
\bibliography{Quasicrystal}

\end{document}